\documentclass[aps, prl, reprint, superscriptaddress, amsfonts, amssymb, amsmath, amssymbf, hyperref]{revtex4-1}
\usepackage{graphicx, dcolumn, bm, hyperref, url}

\newcommand\sci{Science}
\newcommand\ssr{Space Sci. Rev.}
\newcommand\jgr{J. Geophys. Res.}
\newcommand\apjl{Astrophys. J. Lett.}
\newcommand\solphys{Sol. Phys.}
\newcommand\aap{Astron. Astrophys.}
\newcommand\mnras{Mon. Not. R. Astron. Soc.}
\newcommand\grl{Geophys. Rev. Lett.}
\newcommand\jkas{J. Korean Astron. Soc.}
\newcommand\lrsp{Liv. Rev. Sol. Phys.}
\newcommand\natast{Nat. Astron.}

\newcommand\sw{Space Weather}

\newcommand\baas{Bull. American Astron. Soc.}
\newcommand\procspie{Proc. Soc. Photo Opt. Instrum. Eng.}
\newcommand\adsurl{http://adsabs.harvard.edu/abs}
\newcommand\arxurl{http://arxiv.org/abs}

\begin{document}
\title{2D solar wind speeds from 6 to 26 solar radii in solar cycle 24 by using Fourier filtering}

\author{Il-Hyun Cho}
\affiliation{Kyung Hee University, Yongin-si, Gyeonggi-do, 17104, Korea}

\author{Yong-Jae Moon}
\email{moonyj@khu.ac.kr}
\affiliation{Kyung Hee University, Yongin-si, Gyeonggi-do, 17104, Korea}

\author{Valery M. Nakariakov}
\affiliation{Kyung Hee University, Yongin-si, Gyeonggi-do, 17104, Korea}
\affiliation{University of Warwick, Coventry CV4 7AL, UK}

\author{Su-Chan Bong}
\affiliation{Korea Astronomy and Space Science Institute, Yuseong-gu, Daejeon 34055, Korea}
\affiliation{University of Science and Technology, Yuseong-gu, Daejeon 34113, Korea}

\author{Jin-Yi Lee}
\affiliation{Kyung Hee University, Yongin-si, Gyeonggi-do, 17104, Korea}

\author{Donguk Song}
\affiliation{National Institutes of Natural Sciences, Mitaka, Tokyo 181-8588, Japan}

\author{Harim Lee}
\affiliation{Kyung Hee University, Yongin-si, Gyeonggi-do, 17104, Korea}

\author{Kyung-Suk Cho}
\affiliation{Korea Astronomy and Space Science Institute, Yuseong-gu, Daejeon 34055, Korea}
\affiliation{University of Science and Technology, Yuseong-gu, Daejeon 34113, Korea}

\date{\today}

\begin{abstract}
Measurement of the solar wind speed near the Sun is important
for understanding the acceleration mechanism of the solar wind.
In this study, we determine 2D solar wind speeds from 6 to 26 solar radii
by applying Fourier motion filters to \textit{SOHO}/LASCO C3 movies observed from 1999 to 2010.
Our method successfully reproduces the original flow speeds
in the artificially generated data as well as streamer blobs.
We measure 2D solar wind speeds from 1-day to 1-year timescales
and their variation in solar cycle 24.
We find that the solar wind speeds at timescales longer than a month
in the solar maximum period are relatively uniform in the azimuthal direction,
while they are clearly bimodal in the minimum period,
as expected from the \textit{Ulysses} observations and IPS reconstruction.
The bimodal structure appears at around 2006, becomes most distinctive in 2009,
and abruptly disappears in 2010.
The radial evolution of the solar wind speeds resembles the Parker's solar wind solution.
\end{abstract}


\maketitle
\setlength{\parskip}{0.1\baselineskip}

\textit{Introduction.--} The solar wind is a magnetized plasma emanated from the Sun \cite{2007Sci...318.1585S, 2016SSRv..201...55A},
which disturbs the planet atmospheres \cite{2004Sci...305.1933L} and forms the heliosphere \cite{2017NatAs...1E.115D}.
Early in-situ observations of the solar wind were performed mainly in the equatorial plane.
{\it Mariner} 2 observed the continuous and fast solar wind in the range 0.7--1 AU \cite{1966JGR....71.4469N},
firstly confirming that the solar wind is the result of expansion of the hot solar corona \cite{1958ApJ...128..664P}.
{\it Helios} spacecrafts observed important kinetic properties of solar wind in the range 0.3--1 AU \cite[e.g.,][]{1982JGR....87...35M}.
{\it Pioneer mission} explored outer heliosphere and found a large scale radial structure of solar wind \cite{1984ApJ...285..339K}.
The \textit{Ulysses} spacecraft firstly explored the plasma property of the heliosphere in high latitude over $\pm$80$^{\circ}$ \cite{1996ApJ...465L..69S},
providing the direct measurement of the latitudinal structure of the solar wind \cite{1995Sci...268.1005S}.
In-situ observations have been performing continuously by series of near Earth satellites
such as the Interplanetary Monitoring Platform (\textit{IMP}), Advanced Composition Explorer (\textit{ACE}), and \textit{WIND}.

Early spectroscopic measurements in the ultraviolet acquired with rockets and space shuttles provided plasma properties in detail for the extended
corona below 5 solar radii \cite{1999ApJ...511..481C}, revealing that the sonic height of the solar winds is $\sim$2 solar radii
\cite{1982ApJ...254..361W, 1984BAAS...16..531K, 1993ApJ...412..410S}.
It is also found that the line-of-sight (LOS) velocity has multiple components \cite{1995SSRv...72...29K}.
Since then, systematic observations by Ultraviolet Coronagraph Spectrometer (UVCS) \cite{1995SoPh..162..313K} on-board
the Solar and Heliospheric Observatory (\textit{SOHO}) \cite{1995SoPh..162....1D} enables us to study
the plasma property in the extended corona in several aspects;  large scale characteristics \cite{1999A&A...342..592Z, 2000JGR...105.2345S},
its evolution in the solar cycle \cite{2001ApJ...560L.193M, 2005A&A...430..701V}, 
streamers \cite{2015SoPh..290.2043A}, coronal plumes \cite{2007ApJ...658..643R}, and coronal mass ejections (CMEs) \cite{2009ApJ...692.1271L}.
In particular, outflow speeds of protons and heavy ions were calculated by using
Doppler dimming technique \cite{1982SSRv...33...17W, 1987ApJ...315..706N} for various latitude and phase of solar cycle
\cite{1999SSRv...87..311S, 2000SoPh..197..115A, 2001ApJ...560L.193M, 2001ApJ...549L.257M, 2002ApJ...574..477Z, 2003ApJ...597.1145F,
2003ApJ...588..566T, 2005ApJ...635L.185G, 2007A&A...472..299T, 2012A&A...545A...8Z, 2015A&A...577A..34D, 2016A&A...592A.137D, 2017ApJ...846...86B},
revealing how the solar wind speed evolves in the extended corona.

The radio scintillation in the interplanetary space enables to explorer the spatial structure of solar wind speed
in the inner part of the interplanetary space from ground-based observatories in timescales typically longer than a few days.
It is found that the acceleration of solar wind speed in the south pole measured by the interplanetary radio scintillation (IPS) technique
is almost complete at 10 solar radii, suggesting that the acceleration is strongly linked to the coronal heating \cite{1996Natur.379..429G}.
The solar wind speed from 1.5 to 20.5 solar radii for a wide latitudinal range measured by applying the IPS observation on the radio signals from
the Venus Explorer, suggesting that the supply of plasma from closed loops to the solar wind occurs over an extended area \cite{2014ApJ...788..117I}.
The observations with many astrophysical objects combined with a tomographic reconstruction provides the global structure of solar wind
\cite{1998JGR...103.1981K, 2012ApJ...751..128M}.

As the continuous and homogeneous coronagraphic measurements of scattered light by coronal plasma become possible,
the radial evolution of the solar wind speed could be traced in timescales shorter than that of the IPS observations.
The radial speeds tracked by the heights of streamer blobs observed by the Large Angle and Spectroscopic Coronagraph (LASCO) \cite{1995SoPh..162..357B}
on-board the SOHO are well characterized by the Parker's solar wind
that isothermally expanded at a temperature of 1.1 MK and a sonic point near 5 solar radii,
interpreting that the speeds are a passive tracer of the solar wind \cite{1997ApJ...484..472S}.
The radial speeds tracked by slowly evolving CMEs from 2 to 30 solar radii are well fit to a power law
with the exponent lower than 1.0 \cite{2009SoPh..257..351S}, which is similar to the profile of streamer blobs \cite{1997ApJ...484..472S}.
Recently, these structures are traced up to $\sim$50$^{\circ}$ from the Sun \cite{2009ApJ...694.1471S, 2011ApJ...734....7R} in the STEREO HI images
which provides a detailed information on the dynamical evolution of the structures in the interplanetary space far from the Sun.

A few attempts to examine the 2D structure of solar wind speed has been made to understand its dynamical properties in the low corona.
For examples, polar solar wind speeds in the low corona ranging from 2.8 to 10 solar radii, which are calculated where the cross-correlation (e.g., \cite{1999A&A...350..302T, 2002MNRAS.333..969L}) is high,
were found to be a mixture of intermittent slow and fast patches of material \cite{2014ApJ...793...54J}.
Decelerations of the solar wind at $\sim$1.5 solar radii at the poles and $\sim$2 solar radii at the equator were detected by analysing the Doppler dimming
\cite{1982SSRv...33...17W, 1987ApJ...315..706N} from the reconstructed images of the polarized brightness and ultraviolet \cite{2017ApJ...846...86B}.
Here we determine 2D solar wind speeds from 6 to 26 solar radii using coronagraphic observations.

\textit{Data and Method.--} Since 1996, the SOHO/LASCO C3 instrument provides continuous and homogeneous datasets of white-light coronagraphic observations
near the Sun (4--30 solar radii) where most of the dynamical evolution of eruptions and the wind acceleration would be completed.
The coronagraphic white-light image largely show CMEs \cite{2011ApJ...734....7R, 2014ApJ...780...28C, 2017ApJ...840...76H}
associated plasma outflows \cite{2003ApJ...594.1068K, 2013A&A...557A.115K, 2016SoPh..291.3725W, 2017ApJ...841...49C},
shocks \cite{2014ApJ...796L..16L, 2017ApJ...836..246K},
streamer blobs \cite{2009SoPh..258..129S}, and jets \cite{2015ApJ...806...11M}, that are very dynamic.
It also contains less dynamic features such as streamers \cite{2015SoPh..290.2043A}
but these structures are likely to fade in the outer corona and be observed as flowing structures \cite{2016ApJ...828...66D}.
All these dynamic and faint features can be decomposed into a series of movies as a function of speed by applying the Fourier motion filters \cite{2014ApJ...787..124D}, which has been successfully applied to detect the faint inbound motion reflected due to an Alfv\'{e}n surface \cite[e.g.,][]{2009ApJ...700L..39V} in the corona, by keeping the first and third quadrants of the Fourier spectrum of an height-time image.

We use open-access LASCO C3 images from 1999 to 2010 of which the number of images for three consecutive days exceeds 108.
An imaging data cube is obtained by sequentially taking the level 0.5 images that have been rectified to put the solar north at the top of the image,
by using the SolarSoft \cite{64} function \textit{mk$\_$img.pro}
with the keywords \textit{rectified} and \textit{log$\_$scl}.
After removing few bad images, we obtain one movie $I(x, y, t)$ for three days which has a size of 1024$\times$1024$\times n$,
where $n$ is the number of images during three days which typically exceeds 100.

We use a normalized movie $N(x, y, t)$ by taking $(I-I_{MED})/I_{MAD}$ for a given pixel,
where $I_{MED}$ and $I_{MAD}$ represent the median and median absolute deviation of $I(t)$.
We further suppress the intensity of stars and planets into 0$\pm$5$\sigma$.
Their speeds in the movie are similar to the speed of the Earth's revolution ($\sim30$ km s$^{-1}$) so that its
contribution to solar wind speeds is minor.
We transform the movie into the polar coordinate $(r, \theta, t)$ centered on the Sun, with the size of 505$\times$1444$\times n$.
The transformed image for a given time is spatially re-mapped to 128$\times$360 to increase the signal to noise ratio.
Then the movie is temporally interpolated with cadence of 33.75 min (equal to 72 hours / 128), giving a movie $I(\theta, r, t)$ with the size of 128$\times$360$\times$128.
Thus for a given azimuth angle, the size of height-time image is 128$\times$128.

The movie is decomposed into a series of movies as a function of speed by performing the inverse Fourier transform for the filtered spectrum of height-time image, with varying pass band of phase speed for a given azimuth angle as follows,
\begin{equation}
\begin{split}
N(r, \theta_i, t, v_j) = \mathcal{F}^{-1}_{r,t} \{ \mathcal{F}_{r,t} \{I(r, \theta_i, t)\} G(k, w)_{v_j} \},
\end{split}
\end{equation}
where $\theta_i$ ranges 0--360$^{\circ}$ with the sampling size of 1$^\circ$. The phase speed filters that pass the powers around $v_j$ (0, 30, 60, ..., 2010 km s$^{-1}$) are defined as follows,
\begin{equation}
\begin{split}
& G(k, w)_{v_j} = e^{-(v-v_j)^2/2\sigma_v^2} \\
& \times \{ e^{-(k-k_m)^2/2\sigma_k^2} + e^{-(k+k_m)^2/2\sigma_k^2} \} \\
& \times \{ e^{-(\omega-\omega_m)^2/2\sigma_w^2} + e^{-(\omega+\omega_m)^2/2\sigma_w^2} \},
\end{split}
\end{equation}
where $v$, $\sigma_v$, $k_m$, $\sigma_k$, $\omega_m$, $\sigma_\omega$ are $w/k$, 30 km s$^{-1}$, ($5r_{Sun}$)$^{-1}$, ($6r_{Sun}$)$^{-1}$, (5 hours)$^{-1}$, (6 hours)$^{-1}$, respectively.
The operator $\mathcal{F}$ and $\mathcal{F}^{-1}$ are the Fourier transform and inverse Fourier transform, respectively. The latter two terms are low pass filters in the wavenumber and frequency domain, which are symmetric with respect to zero wavenumber or frequency so that the powers around zero slightly decrease when compared to those at the mean values. This may suppress the effect of large scale gradients possibly remained in the movie.
Then we define $N^2(r, \theta, t, v)$ as the speed histogram $P(r, \theta, t, v)$.
Thus the solar wind speed and its standard deviation are defined as follows,

\begin{equation}
\begin{split}
& V(r, \theta, t) = \sum_v{vP(r,\theta,t)}/\sum_v{P(r,\theta,t)}, \\
& \sigma_V(r, \theta, t) = \sqrt{\sum_v{(v-V)^2 P(r,\theta,t)}/\sum_v{P(r,\theta,t)}}.
\end{split}
\end{equation}

We note that original $\sigma_V(r, \theta, t)$ ranges 100--700 km s$^{-1}$ which is rather large compared to $V(r, \theta, t)$, implying that a speed histogram likely has multiple peaks or a broad distribution. It is possibly due to multiple velocity components or low signal-to-noise ratio.
In addition, three projection effects can partly contribute to the errors.
Firstly, our measurements are based on 2D observations that integrate scattered photons on a LOS, which includes different heights from the Sun for a given position. Secondly, there is an offset between the solar equator and the observed center of the Sun, which ranges approximately $\pm$7$^\circ$. Thirdly, an original movie covers 3-days which ranges 40--43$^\circ$ of longitude depending on latitude.
Thus we take the central day from the movie to obtain the map of 1-day median, which includes $\sim$42 samples. If we assume the error is inversely proportional to the square root of the number of samples when averaging, the median absolute deviation (0.6745$\sigma$) becomes 10--72 km s$^{-1}$, which may be acceptable for an application level.

\begin{figure}
\includegraphics[width=85mm]{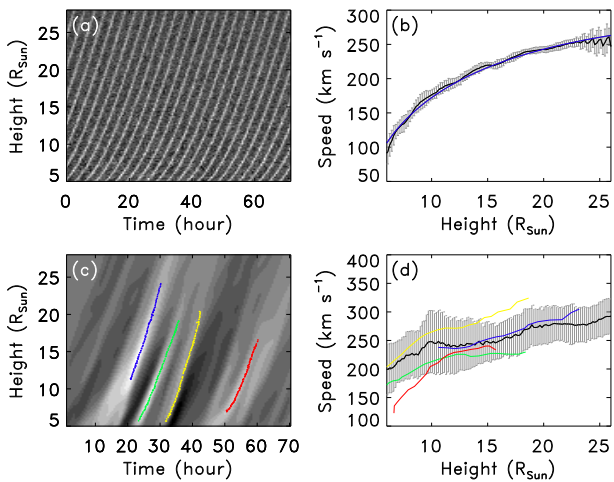}
\caption{Height-time images of artificially generated moving intensity features (white in the panel a) and
streamer blobs observed by LASCO C3 from 2007-Sep-26 to 2007-Sep-28 at azimuth angle equal to 334$^{\circ}$ (white in the panel c),
and comparisons of the estimated speeds by applying Fourier motion filter with true speeds (b and d).
In the right panels, the estimated speeds are indicated by black solid lines.
The true speeds are indicated by blue, green, yellow, and red lines.
The grey vertical bars indicated in the right panels are the median absolute deviations during three days at each height.
}
\label{fig:figure1}
\end{figure}

\begin{figure}
\includegraphics[width=75mm]{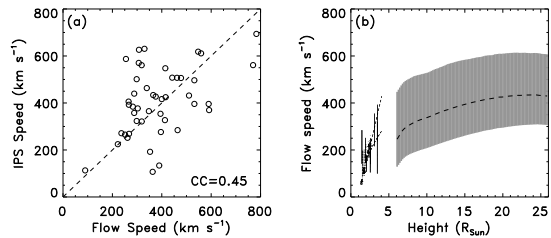}
\caption{Comparisons of the solar wind speeds obtained from LASCO C3 with IPS speeds (a) and
radial speeds of proton given by Doppler dimming technique below 6 solar radii (b).
In panel a, the IPS speeds are interpolated from the projected speeds \cite{65} based on the position and time of our datasets covering whole latitudes.
In panel b, the radial speeds of proton are measured in the extended above coronal holes that have been taken from \cite{2017ApJ...846...86B},
and the estimated electron speed in 6--26 solar radius is taken from north and south poles.
The grey area indicates 1-$\sigma$ interval.}
\label{fig:figure2}
\end{figure}

As shown in Figure \ref{fig:figure1}, the estimated speeds by using the Fourier filtering are well consistent with the true speeds, indicating that the method properly detects solar wind speeds
from pseudo-real and real height-time images.
In Figure \ref{fig:figure1}b, the true speed (blue) is defined as 300 $\sqrt{e^{-(r-4)/15}}$ km s$^{-1}$
to simulate the height-time map (Figure \ref{fig:figure1}a) that is constructed by filing up time-shifted-intensity pulses along the heights.
In Figure \ref{fig:figure1}d, the true speeds (colored lines) are instantaneous speeds calculated from manually determined height-time data in Figure \ref{fig:figure1}c.
The height-time image of streamer blobs (Figure \ref{fig:figure1}c) is filtered by a broadband Gaussian filter (200$\pm$100 km s$^{-1}$)
to improve the visibility of the signal.

It seems that our measurements are roughly consistent with previous ones.
In Figure \ref{fig:figure2}a, we compare $V(r, \theta, t)$ with IPS solar wind speeds \cite{65} which are projected to the sky plane during 1999--2010.
The IPS observations are mostly contributed by the plasma located at the closest position in the LOS
between the observer and a radio source \cite[e.g.,][]{2012ApJ...751..128M, 2013SoPh..285..167S}.
Here we choose the IPS samples with velocity error less than 10 km s$^{-1}$ and interpolated the position and time to those of our datasets.
The correlation coefficient (CC) is 0.45 and its significance level is less than 0.01.
The marginal CC may be due to the difference of characteristic sizes of the local plasma detected by IPS and white-light coronagraph.
In Figure \ref{fig:figure2}b, we plot the $V(r, \theta, t)$ obtained at 90$^{\circ}$ and 270$^{\circ}$ from 1999 to 2010
onto the collection of proton outflow speeds in coronal holes derive from Doppler dimming technique (see \cite{2017ApJ...846...86B}), which may be well connected to the electron speed within 1-$\sigma$ intervals (grey area).

\begin{figure}
\includegraphics[width=85mm]{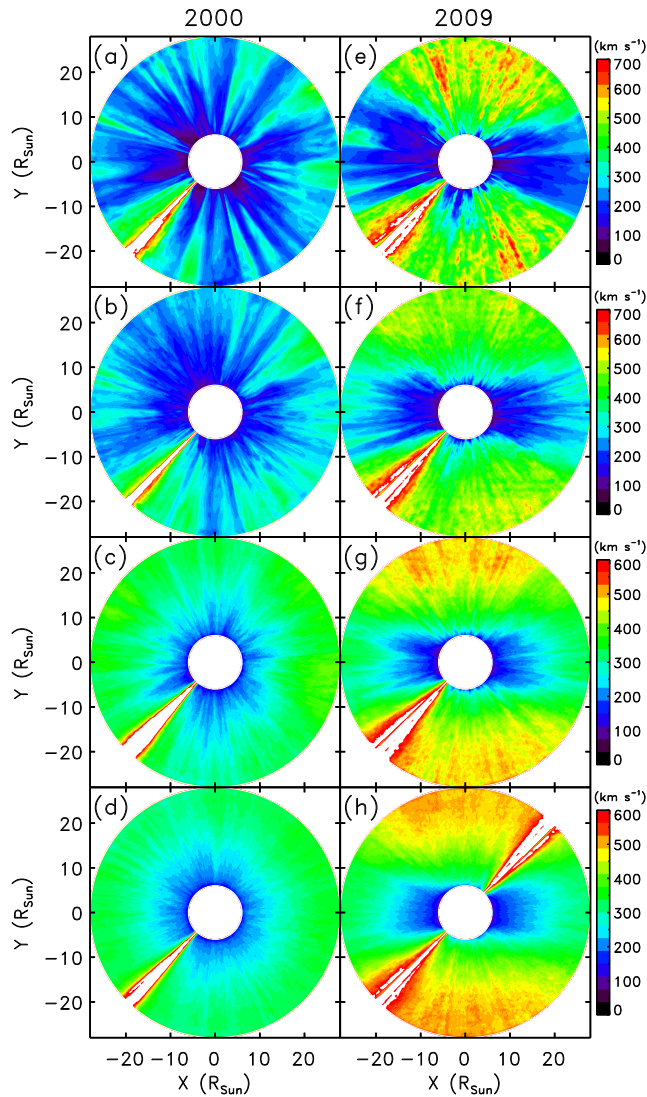}
\caption{Maps of median solar wind speeds in 6--26 solar radii over several time periods (1-day to 1-year) in 2000 (a-d) and 2009 (e-h).
The 1-day maps are constructed by taking the median solar wind speed at each position during 00--24 UT on 2000-Oct-4 (a) and 2009-Mar-7 (e).
The 1-week maps are constructed by taking the median solar wind speed at each position during 2000-Oct-4 12 UT$\pm$3.5 days (b) and 2009-Mar-7 12 UT$\pm$3.5 days (f).
The 1-month maps are constructed from maps of daily solar wind speed in 2000-Jul (c) and 2009-Feb (g).
The 1-year maps are constructed from maps of monthly solar wind speed in 2000 (d) and 2009 (h).}
\label{fig:figure3}
\end{figure}

\textit{Results.--} By using the method, we obtain the maps of the solar wind speed with timescales of half-hour from 1999 to 2010.
From this, we determined the maps of the solar wind speed from 1-day to 1-year.
The spatial structure of the solar wind speed is generally believed to be homogeneous
with respect to the solar latitude in the solar maximum period
because of the frequent appearance of non-polar coronal holes and CMEs in the solar corona.
On the other hand, it shows a bimodal structure in the solar minimum period :
mainly fast winds are detected in the polar regions and slow winds in the equatorial regions \cite{1995GeoRL..22.3301P}.
Figure \ref{fig:figure3} shows maps of median solar wind speeds over several time periods of sampling (1-day to 1-year) in 2000 and 2009.
It is shown that the intrinsic features of the latitudinal distribution of the solar wind speed seem to become evident
as the sampling time period becomes longer.
In 2000, the median solar wind maps over short time periods in Figure \ref{fig:figure3}a and \ref{fig:figure3}b show fast speeds with approximately 400 km s$^{-1}$ for a specific latitude,
possibly contributed by moving features such as CMEs and streamer blobs.
However, the uniform distribution of the solar wind speed with latitude appears in the monthly median.
In 2009, the bimodal structure of solar wind speed becomes apparent in the 1-week median map as in Figure \ref{fig:figure3}f.
Thus we confirm the uniformity and bimodality of the spatial structure of the solar wind speeds in the inner part of the interplanetary space,
which are consistent with previous observations of outer heliosphere \cite{1990SSRv...53..173K, 2000JGR...10510419M} as well as of extended corona below 6 solar radii (e.g. \cite{2017ApJ...846...86B, 2018A&A...612A..84D}).
More precise quantitative measurement can be obtained by future space observations such as
the \textit{Parker Solar Probe} \cite{71, 2016SSRv..204....7F}, Metis coronagraph \cite{2012SPIE.8443E..3HF, 2017SPIE10566E..0LA} on-board Solar Orbiter \cite{2013SPIE.8862E..0EM}, and \textit{ISS} Coronagraph \cite{2017JKAS...50..139C, 2018JKAS...51...27Y}.

\begin{figure*}
\includegraphics[angle=90, width=170mm]{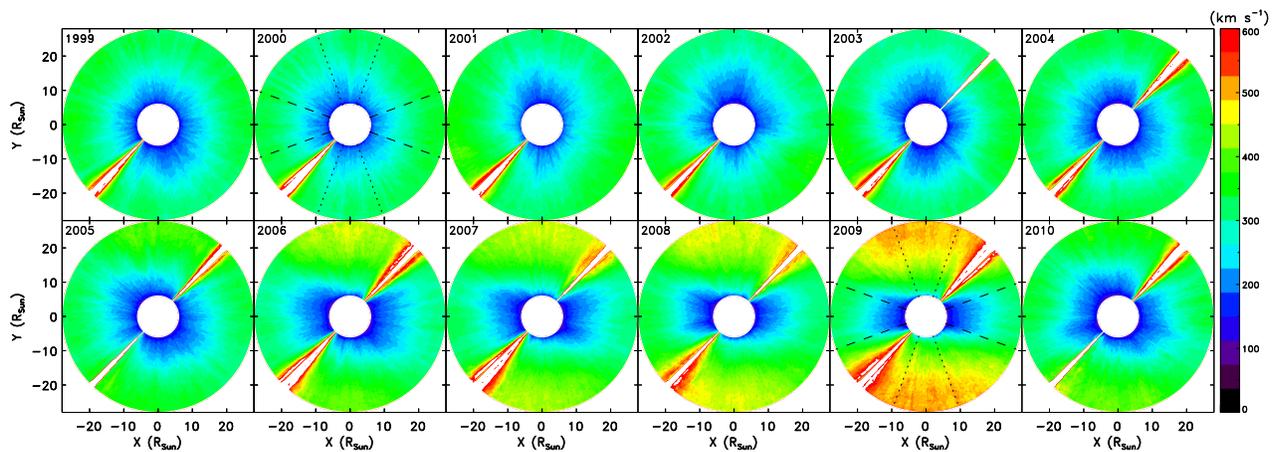}
\caption{Maps of yearly median solar wind speeds from 1999 to 2010.
The two dotted lines in 2000 and 2009 indicate $\pm$20$^{\circ}$ from the north and south poles.
The two dashed lines in 2000 and 2009 indicate $\pm$20$^{\circ}$ from the equator.}
\label{fig:figure4}
\end{figure*}

\begin{figure}
\includegraphics[width=85mm]{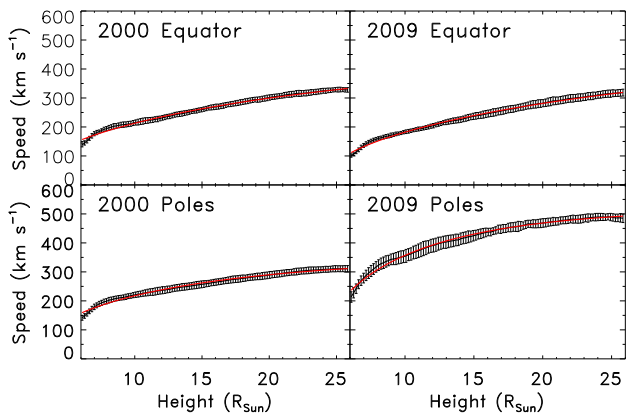}
\caption{Radial profiles of equatorial and polar solar wind speeds in the 1-year map of 2000 and 2009.
The black lines represent the observed speeds averaged over $\pm$20$^{\circ}$ from the equator and poles indicated by dashed and dotted lines in Figure \ref{fig:figure4},
respectively.  The vertical bars are the corresponding standard deviations.
The red lines represent the model speeds obtained from the least squares fit by the function defined as $v_0\sqrt{ 1-e^{(r-r_1)/r_0} }$.}
\label{fig:figure5}
\end{figure}

The uniform and bimodal structures are more clear in the yearly maps of the solar wind speed as seen in Figure \ref{fig:figure4}.
The uniform latitudinal distribution is observed at the starting year and maintain during the solar maximum period (1999--2004).
The latitudinal structure of the solar wind speed does not seem to change dramatically during the maximum period.
It is evident that the bimodal structure started to appear in 2005 and grew until 2009 which corresponds to the solar minimum.
The structure in 2009 is the most apparent. The polar solar wind speed seems to increase as the solar activity goes to the activity minimum
which is similar to the result given by \cite{2010JGRA..115.4102T}.
The bimodal structure disappears and becomes uniform around 2010 which corresponds to the beginning of the new solar cycle 24.

In Figure \ref{fig:figure5}, radial profiles of solar wind speeds near the solar poles and equator in the 1-year map in 2000 and 2009 are presented.
The profiles are well matched with the function $v_0\sqrt{ 1-e^{-(r-r_1)/r_0} }$,
which describes a rapid acceleration until $r\sim r_1$ and approaches the asymptotic speed $v_0$ at $r - r_1\gg r_0$ \cite{1997ApJ...484..472S},
resembling the Parker's solar wind solution.
In other words, the acceleration is rapid in the initial stage and exponentially decreases with height.
Our study is consistent with the previous indirect observation that
the polar solar wind is mainly accelerated below 10 solar radii and then continues at a nearly constant speed \cite{1996Natur.379..429G}.

The 2D solar wind speeds in various timescales from 1-day to 11-year determined in our study could be compared with measurements in the heliosphere \cite{2016SSRv..201...55A, 2017AGUFMSH21A2648B, 80}, and possibly in astrospheres \cite{2004LRSP....1....2W}. The propagation of CMEs is affected by the properties of CME itself as well as the background wind speed \cite[e.g.,][]{2001JGR...10629207G, 2006SoPh..235..345M, 2007JGRA..112.5104K, 2007A&A...472..937V, 2010A&A...512A..43V, 2011ApJ...743..101T, 2014JGRA..119.7120J}. Thus we hope that our results would be used to improve the accuracies of CME arrival times \cite[e.g.,][]{2014SpWea..12..448Z, 2018ApJ...855..109L}. The speeds also provide constraints on the solar wind models \cite[e.g.,][]{1966PhRvL..16..628S, 2011A&A...529A.148L, 2012SSRv..172..145C, 2014ApJ...788...43U, 2017ApJ...838...89P}, which would improve our understanding of the dynamical properties of the solar wind.

\begin{acknowledgments}
We appreciate the constructive comments from anonymous referee which improved the manuscript.
The \textit{SOHO}/LASCO data used here are produced by a consortium of the Naval Research Laboratory (USA),
Max-Planck-Institut for Sonnensystemforschung (Germany),
Laboratoire d'Astrophysique Marseille (France),
and the University of Birmingham (UK).
\textit{SOHO} is a project of international cooperation between ESA and NASA.
This research is supported by the Korea Astronomy and Space Science Institute under the R\&D program,
Development of a Solar Coronagraph on International Space Station (Project No. 2017-1-851-00),
supervised by the Ministry of Science, ICT
and the BK21 plus program through the National Research Foundation (NRF) funded by the Ministry of Education of Korea.
Y.J.M. acknowledges the support from Basic Science Research Program through the NRF funded by Ministry of Korea (NRF-2016R1A2B4013131)
and NRF of Korea Grant funded by the Korean Government (NRF-2013M1A3A3A02042232).
V.M.N is supported by the STFC consolidated grant ST/P000320/1.
J.Y.L. is supported by Basic Science Research Program though the NRF funded by the Ministry of Education of Korea (NRF-2016R1A6A3A11932534).
\end{acknowledgments}

\end{document}